\documentclass[aps,prl,twocolumn,showpacs,amsmath,amssymb,superscriptaddress,citeautoscript]{revtex4-1}
\usepackage{graphicx}
\usepackage{dcolumn}

\begin{document}

\title{Andreev reflection spectroscopy on Bi$_{2}$X$_{3}$ (X = Se, Te) topological insulators: Implications for the \textit{c}-axis superconducting proximity effect}

\author{C. R. Granstrom}
\affiliation{Department of Physics, University of Toronto, 60 St. George St., Toronto, Ontario M5S 1A7, Canada}

\author{I. Fridman}
\affiliation{Department of Physics, University of Toronto, 60 St. George St., Toronto, Ontario M5S 1A7, Canada}

\author{H.-C. Lei}
\altaffiliation{Present address: Department of Physics, Renmin University, Beijing 100872, China}
\affiliation{Condensed Matter Physics \& Materials Science Department, Brookhaven National Laboratory, NY 11973, USA}

\author{C. Petrovic}
\affiliation{Condensed Matter Physics \& Materials Science Department, Brookhaven National Laboratory, NY 11973, USA}

\author{J. Y. T. Wei}
\affiliation{Department of Physics, University of Toronto, 60 St. George St., Toronto, Ontario M5S 1A7, Canada}
\affiliation{Canadian Institute for Advanced Research, Toronto, Ontario M5G 1Z8, Canada}



\begin{abstract}

Using Andreev reflection (AR) as an experimental gauge of the superconducting proximity effect (PE), we assess the topological purity of the superconductivity that is induced by the $c$-axis PE between an $s$-wave superconductor and the topological insulators Bi$_{2}$X$_{3}$ (X=Se,Te).  Point-contact AR spectroscopy is performed with Nb tips on Bi$_{2}$X$_{3}$ single crystals at 4.2 K.  Scanning tunneling spectroscopy is also used, to locate the Fermi level $E_F$ relative to the Dirac point in the crystals. The AR data is analyzed with Blonder-Tinkham-Klapwijk theory, taking into account tip-induced spin-orbit coupling, Fermi-surface mismatch, and the co-presence of bulk band and topological surface states at $E_F$. Our results indicate that the superconductivity that can be proximity-induced into Bi$_{2}$X$_{3}$ is predominantly non-topological.
\end{abstract}

\maketitle

Three-dimensional topological insulators (TIs) are a novel class of materials where strong spin-orbit coupling causes bulk band inversion, creating surface states with a Dirac dispersion as well as helical spin polarization that protects them against back-scattering \cite{Hasan2010,Qi2011,Ando2013}. Ideally, TIs are bulk insulators with high surface mobility from these topological surface states (TSS).  Realistically, intrinsic doping from antisite and vacancy defects  \cite{Ando2013} tends to shift the Fermi level out of the bulk band gap, allowing non-topological bulk band states (BBS) with finite $\hat{k}_z$-dispersion to also carry current.  For the bismuth-chalcogenide TIs, such non-ideal behavior is widely seen in single crystals and thin films of Bi$_{2}$X$_{3}$ (X = Se,Te).

It was theoretically proposed that the TSS on an ideal TI can, through the proximity effect (PE) with a spin-singlet superconductor (SC), be used to generate topological superconductivity and thereby Majorana fermion states \cite{*[{ }] [{, and references therein.}] Beenakker2013,Zutic2018}. Following this proposal, there have been numerous experimental studies of the superconducting PE on Bi$_{2}$X$_{3}$ samples, most of which show non-ideal TI behavior \cite{Sacepe2011,Zhang2011,Veldhorst2012,Wang2012,Williams2012,Xu2014,Molenaar2014,Kurter2014,Xu2015,Sun2016,Li2017,Shvetsov2017,Dai2017}.  These experiments are primarily based on \textit{s}-wave pairing in two types of heterostructures (see Figure S1 in supplement): TI/SC, where the PE is primarily along the \textit{c}-axis; and SC/TI/SC, where the PE is in the \textit{a}-\textit{b} plane.  In this work, we focus on the \textit{c}-axis PE in the TI/SC geometry \cite{Chiu2016} and ask the central question: When BBS and TSS are both present at the Fermi level of a non-ideal TI such as Bi$_{2}$X$_{3}$, how topological is the superconductivity that can be proximity-induced across the TI's \textit{c}-axis interface? 

At a metal/SC interface, Andreev reflection (AR) is the process that converts electrons into Cooper pairs through retro-reflection of holes, and is considered to be the underlying mechanism for the PE \cite{Pannetier2000,Klapwijk2004}.  For conventional metals, AR has been extensively studied and is quite well understood \cite{Blonder1982,Blonder1983,Deutscher2005,Nadgorny2011}. For ferromagnetic metals, AR between spin-polarized electrons and spin-singlet pairs is suppressed, and thus the PE between them is also suppressed \cite{Jong1995,Upadhyay1998,Soulen1998,Ji2001,Zutic2004,Buzdin2005,Nadgorny2011,Turel2011,Chen2012}. The special attributes of TIs may also affect AR and thus the PE, since a Dirac dispersion can modify AR by favoring specular- over retro-reflection \cite{Beenakker2006,Linder2008}, and Rashba spin-orbit coupling can weaken AR also by upsetting retro-reflection \cite{Yokoyama2006,Wu2010,Kononov2013,Hoegl2015}.  A more important issue involves the suppression of AR by Fermi-surface mismatch across the TI/SC interface \cite{Blonder1983,Daghero2011,Yilmaz2014}.  Namely, since the TSS have no $\hat{k}_z$-dispersion, how well can they actually sustain AR and thus proximity-couple with a SC across the TI's \textit{c}-axis interface?

To address these questions, we perform point-contact spectroscopy (PCS) at 4.2 K using Nb tips on \textit{c}-axis faces of Bi$_{2}$X$_{3}$ single crystals.  Scanning tunneling spectroscopy (STS) is also used, to determine the location of the Fermi level in the crystals relative to the Dirac point.  The PCS data show robust AR characteristics and are analyzed with the Blonder-Tinkham-Klapwijk (BTK) theory, taking into account tip-induced spin-orbit coupling, Fermi-surface mismatch, and the co-presence of BBS and TSS at the Fermi level. Spectral analysis based on realistic band structures indicates that the $c$-axis AR is dominated by the BBS over the TSS, implying that the superconductivity that can be proximity-induced into Bi$_{2}$X$_{3}$ is predominantly non-topological.

\begin{figure}[th]
\includegraphics[width=0.42\textwidth]{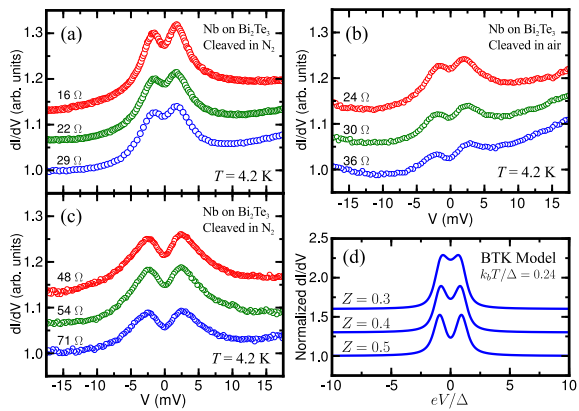}
\caption{\label{fig:NbBi2Te3} 
PCS data using Nb tips on Bi$_{2}$Te$_{3}$ crystals at 4.2 K for varying junction impedances and cleaving environments. To facilitate comparison, each $dI/dV$ spectrum is divided by its above-gap value at -15 mV and staggered vertically. All spectra show a double-hump AR structure. (a) and (b) show data taken on N$_2$- and air-cleaved crystals, respectively. (c) shows data for a N$_2$-cleaved crystal, taken at higher impedances than in (a). In (a), the zero-bias conductance (ZBC) increases monotonically as junction impedance increases, contrary to generic spectral dependence on interfacial barrier strength $Z$ predicted by the BTK model \cite{Blonder1982} in (d).}
\end{figure}

PCS is a well-established technique for studying AR \cite{Nadgorny2011}.  It has the advantage of inherently small junctions that are amenable to BTK modeling and can be checked for data reproducibility at different points on the sample. Our PCS measurements were made at 4.2 K using a scanning tunneling microscope (STM).  Nb tips were used because Nb is a well-studied superconductor and devoid of multigap pairing effects that may complicate data analysis.  The STM technique enabled the tip approach to be gentle, and the junction impedance to be varied by piezos after tip-sample contact is made. STS measurements were also made, using either Nb tips at 4.2 K to characterize the SC tip, or Pt-Ir tips down to 300 mK to characterize the TI crystals. The Bi$_{2}$X$_{3}$ single crystals were $n$-doped as a result of natural defects that commonly occur in bismuth chalcogenides \cite{Ando2013}, and were cleaved before cooldown and measurement. Details of our crystal growth, surface preparation, and measurement technique are given in the supplement.

Figure \ref{fig:NbBi2Te3} plots PCS data taken with Nb tips on Bi$_{2}$Te$_{3}$ crystals at 4.2 K. To compare all data in each plot on the same scale, each differential conductance $dI/dV$ spectrum is divided by its above-gap value at $- 15$ mV, and staggered vertically for clarity. Panels (a) and (b) show data taken on N$_2$- and air-cleaved crystals, respectively.  Panel (c) shows data for a N$_2$-cleaved crystal, taken at higher impedances than in (a).  All spectra have a double-hump structure, characteristic of subgap enhancement due to AR, as well as a background asymmetry favoring positive voltage. The humps are separated by $\sim$ 3.5 - 5 mV, comparable to the gap-edge separation seen by our STS data using Nb tips on Ag films (Figure S2). The subgap enhancement is generally stronger in (a) and (c) than in (b), consistent with the crystal surface being cleaner for N$_2$-cleaving than air-cleaving.  The stronger subgap enhancement of (a) relative to (c) implies stronger AR for lower junction impedance, consistent with the BTK theory.  However, it is notable that the zero-bias conductance (ZBC) in (a) decreases with lower junction impedance, contrary to generic spectral behavior predicted by the BTK model in panel (d).  Estimates of the junction size using the Wexler formula show all junctions to be ballistic \footnote{The Wexler formula is $R = \left(2 h\right)/\left(ek_{F}a\right)^2+\rho/\left(2a\right)$, where $R$ is point-contact impedance, $h$ is the Planck constant, $a$ is contact radius, $\rho$ is electrical resistivity, $e$ is electron charge, and $k_{F}$ is Fermi wavevector \cite{Wexler1966}. For our $R$ of $16-71\,\Omega$, we use $\rho \approx 5 \times10^{-5}$ $\Omega\,\mathrm{\,cm}$ as measured in our crystals and $k_{F} \approx $ 0.5 nm$^{-1}$ as reported for similar crystals \cite{Koehler1976,Qu2010} to estimate $a\approx 56-122$ nm, which is smaller than the mean free path $\ell=150$ nm \cite{Qu2010}}, indicating that diffusive effects are not the source of this non-BTK behavior at low impedance.

\begin{figure}[ht]
\includegraphics[width=0.42\textwidth]{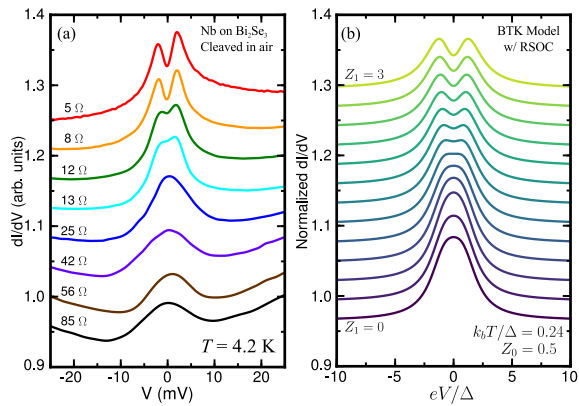}
\caption{\label{fig:NbBi2Se3} 
(a) PCS data using a Nb tip on an air-cleaved Bi$_{2}$Se$_{3}$ crystal at 4.2 K.  To facilitate comparison, each $dI/dV$ spectrum is divided by its above-gap value at -25 mV and staggered vertically. A single- to double-hump evolution is observed as junction impedance decreases from 85 to 5 $\Omega$, with the ZBC evolving non-monotonically. (b) Spectral simulations using a BTK model that includes interfacial Rashba spin-orbit coupling (RSOC) \cite{Wu2010}, where $Z_0$ is a fixed interfacial barrier strength and $Z_1$ is a variable interfacial RSOC strength. As $Z_1$ increases from 0 to 3, the ZBC non-monotonically increases then decreases, similar to (a). A fixed broadening parameter $\Gamma$ of 1.4 meV was used in (b).}
\end{figure}

Figure \ref{fig:NbBi2Se3}(a) shows PCS data taken with a Nb tip on an air-cleaved Bi$_{2}$Se$_{3}$ crystal at 4.2 K.  To facilitate comparison, each $dI/dV$ spectrum is divided by its above-gap value at $-25$ mV and staggered vertically. Here, a single- to double-hump evolution is observed as the junction impedance decreases, with the ZBC evolving non-monotonically. This spectral behavior tends to be seen for lower junction impedances, and is very similar to the low-impedance Bi$_{2}$Te$_{3}$ data in Figure \ref{fig:NbBi2Te3}(a), where the ZBC also evolves in a non-BTK manner.

The non-BTK behavior seen at low junction impedance in Figures \ref{fig:NbBi2Te3}(a) and \ref{fig:NbBi2Se3}(a) can be explained in terms of interfacial Rashba spin-orbit coupling (RSOC), which is known to affect AR at low energies \cite{Yokoyama2006,Wu2010,Kononov2013,Hoegl2015}.  For Bi$_{2}$Se$_{3}$, RSOC has been observed on crystal surfaces and attributed to adsorbate-induced band bending that produces a local electric field \cite{Benia2011}. In our point-contact junctions, band bending can likewise be induced by tip-sample contact due to the large Fermi-energy difference between metallic Nb and semimetallic Bi$_{2}$Se$_{3}$.  Since band bending is enhanced by junction transparency, the tip-induced RSOC would increase as the junction impedance decreases. This RSOC-impedance relationship is consistent with spectral simulations made with a BTK model that includes interfacial RSOC \cite{Wu2010}, shown in Figure \ref{fig:NbBi2Se3}(b). Here, as the RSOC strength $Z_1$ increases from 0 to 3, the ZBC non-monotonically increases then decreases, similar to our data in Figure \ref{fig:NbBi2Se3}(a).

\begin{figure}
\includegraphics[width=0.39\textwidth]{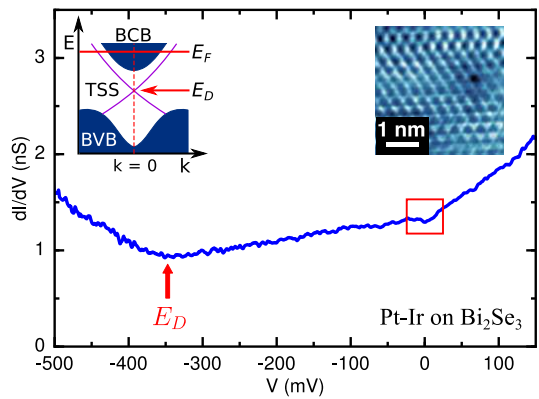}
\caption{\label{fig:Bi2Se3SpecTop} STS data taken on a Bi$_{2}$Se$_{3}$ crystal with a Pt-Ir tip at 4.2 K. The Fermi level is in the bulk conduction band (BCB) and the Dirac point is at -350 mV, as described in the text and shown in the schematic band structure (left inset).  Red box indicates the range over which the PCS data in Figure \ref{fig:NbBi2Se3} were taken.  Right inset shows an STM image of the crystal.}
\end{figure}

We note that a broadening parameter $\Gamma$ of 1.4 meV was needed in Figure \ref{fig:NbBi2Se3}(b) to produce ZBC heights similar to what we observe in Figure \ref{fig:NbBi2Se3}(a).  We can qualitatively explain this non-ideal behavior in terms of spectral broadening from inelastic scattering at the tip-sample interface, which is generally higher in Nb compared to other elemental superconductors, and is known to increase with increasing junction impedance \cite{Naidyuk1996,Slobodzian2002}. Additionally, large spectral broadening has also been observed in recent high-impedance PCS measurements using Nb tips on bulk-insulating (Bi$_{1-x}$Sb$_x$)$_2$Te$_3$ films \cite{Borisov2016}. Further theoretical study is needed to better explain this non-BTK behavior seen at high junction impedance.  Regardless, the subgap conductance enhancements observed in all of our PCS data on Bi$_{2}$X$_{3}$ are robust signatures of AR.

\begin{figure}
\includegraphics[width=0.39\textwidth]{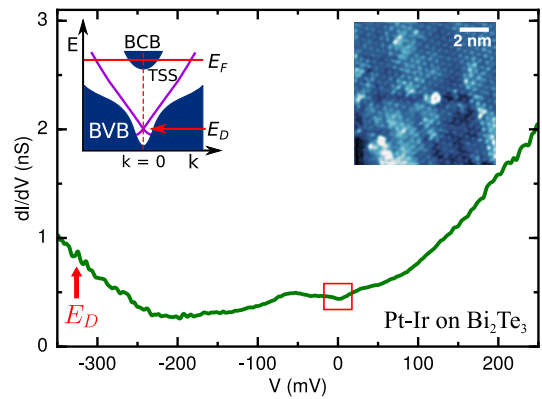}
\caption{\label{fig:Bi2Te3SpecTop} STS data taken on a Bi$_{2}$Te$_{3}$ crystal with a Pt-Ir tip at 300 mK.  The Fermi level is in the BCB and the Dirac point is at -330 mV, as described in the text and shown in the schematic band structure (left inset).  Red box indicates the range over which the PCS data in Figure \ref{fig:NbBi2Te3} were taken.  Right inset shows an STM image of the crystal.}
\end{figure}

To determine the location of the Fermi level $E_{F}$ relative to the Dirac point $E_{D}$ in the Bi$_{2}$X$_{3}$ crystals, we examine cryogenic STS data taken with Pt-Ir tips.  Figures \ref{fig:Bi2Se3SpecTop} and \ref{fig:Bi2Te3SpecTop} show $dI/dV$ spectra taken on atomically-smooth Bi$_{2}$X$_{3}$ surfaces, which are imaged by STM as shown in the insets.  First, it is worth noting that the spectral asymmetry of the PCS data is also seen in the STS data, within the voltage range shown by the red box in each figure.  Next, a comparison of our STS data for Bi$_{2}$Se$_{3}$ (Figure \ref{fig:Bi2Se3SpecTop}) with data taken by angle-resolved photoemission spectroscopy (ARPES) on similar crystals \cite{Xia2009} indicates that our crystals' $E_{F}$ lies in the BCB, $\sim 350$ meV above $E_{D}$.  For Bi$_{2}$Te$_{3}$ however, ARPES has shown \cite{Alpichshev2010} that bulk bands cross $E_{D}$, meaning the minimum in the STS spectrum does not coincide with $E_{D}$ as in the case of Bi$_{2}$Se$_{3}$.  Nevertheless, using similar spectral analysis as Ref.  \cite{Alpichshev2010}, we infer that $E_{F}$ in our Bi$_{2}$Te$_{3}$ crystals also lies in the BCB, $\sim330$ meV above $E_{D}$.

The fact that $E_{F}$ crosses the BCB in our Bi$_{2}$X$_{3}$ crystals indicates that, in addition to TSS, BBS can also take part in AR with the Nb tip.  This raises a natural question: to what extent do our observed AR characteristics involve TSS versus BBS?  We can qualitatively address this question with the BTK theory by considering the geometry of our TI/SC junction.  In the standard BTK model, junction opacity is given by $Z$ = $\sqrt{Z_b^2 + (1-r_v)^{2}/4r_v}$, where $Z_b$ represents interfacial barrier strength and $r_v$ = $v^N_F / v^{SC}_F$ is the ratio of longitudinal Fermi velocities across the N/SC interface. For the TSS, vanishing $\hat{k}_z$-dispersion implies severe $v_F$-mismatch ($r_v \ll 1$) with Nb along the $c$-axis, thus suppressing AR ($Z \gg 1$) even for a perfectly metallic interface ($Z_b = 0$). Such a large $Z$ is inconsistent with our measured subgap enhancement. The BBS, however, have substantial $\hat{k}_z$-dispersion, so their $v_F$-matching with Nb is significantly better, allowing AR to occur. Comparing the $Z$ values estimated with this model for the TSS/Nb and BBS/Nb scenarios \footnote{Using measured $v_F$ values from Table S1 in the supplement and letting $Z_b=0$, $Z=0.1$ and 0.5 for the BBS in Bi$_2$Se$_3$ and Bi$_2$Te$_3$, respectively, while $Z=\infty$ for the TSS in both Bi$_2$Se$_3$ and Bi$_2$Te$_3$} shows that our AR data is better explained by the latter.

\begin{figure}[th]
\includegraphics[width=0.41\textwidth]{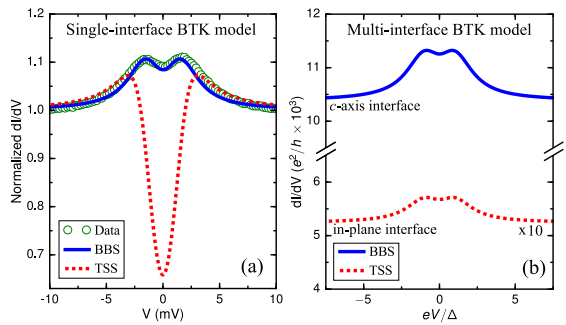}
\caption{\label{fig:BTK_models} 
Spectral simulations using two BTK models of a Bi$_2$X$_3$/Nb junction. (a) Normalized $dI/dV$ spectra for a BTK model that accounts for dependence of junction transmission on injection angle. The TSS/Nb spectrum (dashed curve) is gapped, indicating AR suppression, while the BBS/Nb spectrum (solid curve) shows a double-hump AR structure.  For comparison, a normalized experimental $dI/dV$ spectrum from Figure \ref{fig:NbBi2Te3}(a) is also plotted (open circles). (b) $dI/dV$ spectra (in units of $e^2/ h$) for a junction with $c$-axis and in-plane interfaces, where spectral weighting for each interface is determined by the corresponding number of Landauer channels. The BBS/Nb channels have a stronger spectral weight than the TSS/Nb channels.}
\end{figure}

The above analysis can be refined by using extensions of the BTK model \cite{Tanaka1995,Mortensen1999,Daghero2011} that account for the dependence of junction transmission on injection angle $\theta$, as detailed in the supplement.  Essentially, conservation of momentum transverse to the junction normal introduces a Fermi wavevector-matching term $k^N_F / k^{SC}_F$ and an angle-dependent barrier term $Z_{b}/\cos\theta$ that both enter into $Z$.  Spectral simulations based on the model from Ref. \cite{Mortensen1999} are plotted in Figure \ref{fig:BTK_models}(a) to compare the TSS/Nb and BBS/Nb scenarios.  The TSS/Nb spectrum appears gapped, indicating suppression of AR, while the BBS/Nb spectrum shows a double-hump AR structure similar to our background-normalized data (see supplement for details). We note that subgap suppression similar to the TSS curve in Figure \ref{fig:BTK_models}(a) was recently observed in high-impedance PC junctions on bulk-insulating (Bi$_{1-x}$Sb$_x$)$_2$Te$_3$ films \cite{Borisov2016}, and was attributed to the helical spin-polarization of the TSS.

\begin{figure}
\includegraphics[width=0.32\textwidth]{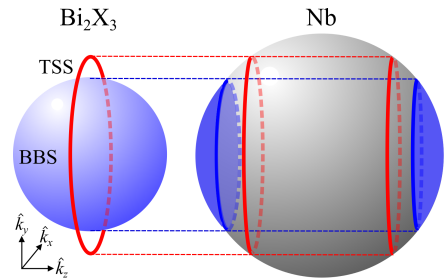}
\caption{\label{fig:fermi_surf_proj} Schematic Fermi-surface projections between Bi$_{2}$X$_{3}$ and Nb, corresponding to conservation of Fermi wavevector transverse to the junction normal ($\hat{k}_z$).  The BBS Fermi surface (left sphere) projects as two domes onto the Nb Fermi surface (right sphere), whereas the TSS Fermi surface (left circle) projects as two circles onto the Nb Fermi surface. The phase space of states available for AR is clearly larger for BBS than for TSS.}
\end{figure}

To visualize why $c$-axis AR is suppressed for TSS but robust for BBS, we examine the Fermi-surface projections between Bi$_{2}$X$_{3}$ and Nb along $\hat{k}_z$.  As illustrated in Figure \ref{fig:fermi_surf_proj}, the BBS Fermi sphere projects as two domes onto the Nb Fermi sphere, whereas the TSS Fermi surface is a circle that projects as two circles onto the Nb Fermi sphere.  Three key facts are worth noting. First, since these projections correspond to allowed interfacial scattering processes and thus include states that allow AR, the phase space for AR is clearly larger for BBS than for TSS.  Second, although TSS and BBS are similarly $k_F$-matched with Nb, the $v_F$-matching for TSS/Nb is invariably poor because TSS have vanishing dispersion along $\hat{k}_z$, and thus $Z\gg1$.  Third, since TSS/Nb transmission corresponds to $\theta=\pi /2 $, it is suppressed by the $Z_{b}/\cos\theta$ term for any non-zero $Z_b$.  All three facts indicate that TSS contribute negligibly to $c$-axis AR when BBS are also present.

One possible way that TSS can contribute to AR is if the junction has regions that are not normal to the $c$-axis of Bi$_{2}$X$_{3}$.  Since the TSS has finite dispersion in the $\hat{k}_x$-$\hat{k}_y$ plane, these non-normal regions may allow better $v_F$-matching with Nb and thus small enough $Z$ for AR to occur.  To consider this possibility, we assume that a Bi$_{2}$X$_{3}$/Nb point-contact has both $c$-axis and in-plane interfaces with area $\pi a^2$ and circumference $2 \pi a $ respectively, where $a$ is the contact radius.  Applying the BTK model from Figure \ref{fig:BTK_models}(a) to both interfaces, we find that AR can now occur for TSS/Nb channels via the in-plane interface, but much more weakly than the BBS/Nb channels via the $c$-axis interface. For this multi-interface model, Figure \ref{fig:BTK_models}(b) plots the $dI/dV$ spectra in units of $e^2/ h$, where the spectral weighting is determined by the corresponding number of Landauer channels for each interface (see supplement), showing double-hump AR characteristics for both the TSS/Nb and BBS/Nb cases.  Once again, the larger AR phase space for the BBS/Nb channels gives them stronger spectral weight compared to the TSS/Nb channels. 

Finally, the inherently larger AR phase space for BBS versus TSS has an important implication for the superconducting PE in Bi$_{2}$X$_{3}$/Nb junctions.  Since AR is the underlying mechanism for the PE, this phase-space disparity again favors BBS over TSS.  On the reasonable assumption that the PE induces topological superconductivity on TSS but non-topological superconductivity on BBS, the greater spectral weighting of BBS over TSS in our observed AR implies that the superconductivity that can be proximitized by Nb across the $c$-axis interface of Bi$_{2}$X$_{3}$ has a greater non-topological component than a topological one. In order to suppress the non-topological component, one would need to suppress BBS by shifting the $E_F$ of Bi$_{2}$X$_{3}$ to within the bulk band gap, for example by compensation doping, electrostatic gating, or photo-illumination \cite{Ando2013,Chang2015,Chen2010,Frantzeskakis2015}. Recent PE and PCS experiments that use bulk-insulating samples \cite{Borisov2016,Stehno2017,Sun2016} and study dependence on bulk-conduction \cite{Xu2015,Stehno2016,Jauregui2017,Shvetsov2017,Wiedenmann2017} have made significant progress in this regard.  However, as $E_F$ moves close enough to $E_D$, specular AR may become dominant enough to degrade conventional AR \cite{Beenakker2006,Linder2008} and thereby the PE \cite{Komatsu2012}.  More studies are necessary to elucidate this complex phenomenon, in order to generate predominantly topological superconductivity via $c$-axis proximity coupling between an \textit{s}-wave superconductor and a topological insulator.

Work was supported by NSERC, CFI-OIT, and the Canadian Institute for Advanced Research. Work at Brookhaven National Laboratory was supported by the U.S. DOE-BES, Division of Materials Science and Engineering, under contract No. DESC00112704.

\bibliography{./AR_Bi2X3_PE}

\end{document}



\title{Supplemental material for \\ ``Andreev reflection spectroscopy on Bi$_{2}$X$_{3}$ (X = Se, Te) topological insulators: Implications for the \textit{c}-axis superconducting proximity effect''}

\author{C. R. Granstrom}
\affiliation{Department of Physics, University of Toronto, 60 St. George St., Toronto, Ontario M5S 1A7, Canada}

\author{I. Fridman}
\affiliation{Department of Physics, University of Toronto, 60 St. George St., Toronto, Ontario M5S 1A7, Canada}

\author{H.-C. Lei}
\altaffiliation{Present address: Department of Physics, Renmin University, Beijing 100872, China}
\affiliation{Condensed Matter Physics \& Materials Science Department, Brookhaven National Laboratory, NY 11973, USA}

\author{C. Petrovic}
\affiliation{Condensed Matter Physics \& Materials Science Department, Brookhaven National Laboratory, NY 11973, USA}

\author{J. Y. T. Wei}
\affiliation{Department of Physics, University of Toronto, 60 St. George St., Toronto, Ontario M5S 1A7, Canada}
\affiliation{Canadian Institute for Advanced Research, Toronto, Ontario M5G 1Z8, Canada}

\renewcommand{\thefigure}{S\arabic{figure}}

\setcounter{figure}{0}

\renewcommand{\thetable}{S\arabic{table}} 
 
\setcounter{table}{0}

\pacs{74.45.+c, 73.20.At, 73.63.Rt}

\maketitle

\section{S1. Geometries used in TI/SC PE experiments}

\begin{figure}[ht]
\includegraphics[width=0.4\textwidth]{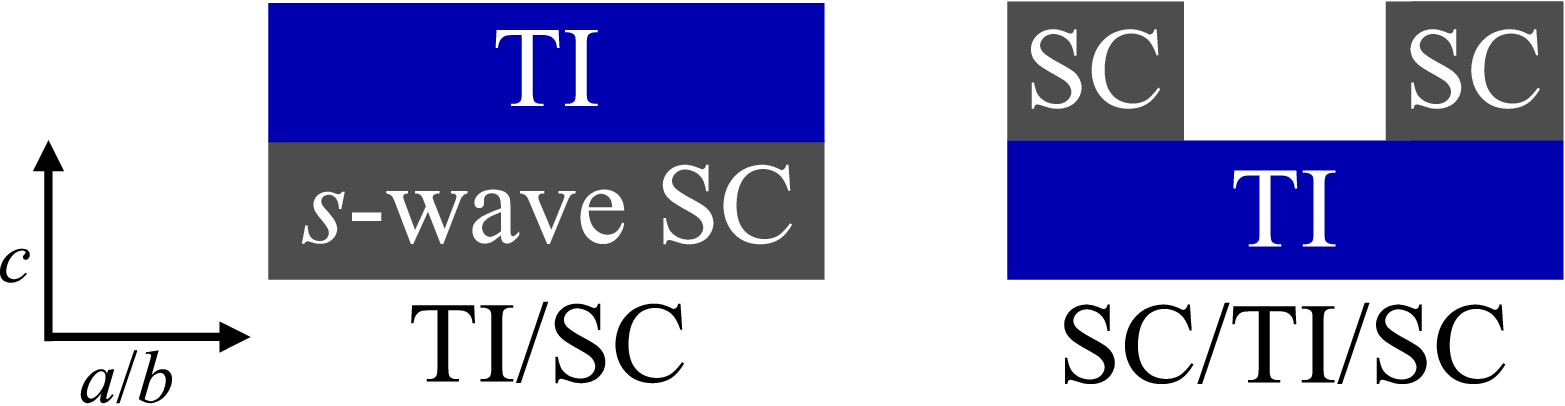}
\caption{\label{fig:geometries} Two types of heterostructures used in proximity-effect experiments between Bi$_{2}$X$_{3}$ TIs and $s$-wave SCs.}
\end{figure}

Experimental studies of the PE on Bi$_{2}$X$_{3}$ samples have primarily been based on \textit{s}-wave pairing in two types of heterostructures (Figure \ref{fig:geometries}): TI/SC, where the PE is primarily along the \textit{c}-axis; and SC/TI/SC, where the PE is in the \textit{a}-\textit{b} plane.

\section{S2. Experimental Details}

\begin{figure}[ht]
\includegraphics[width=0.35\textwidth]{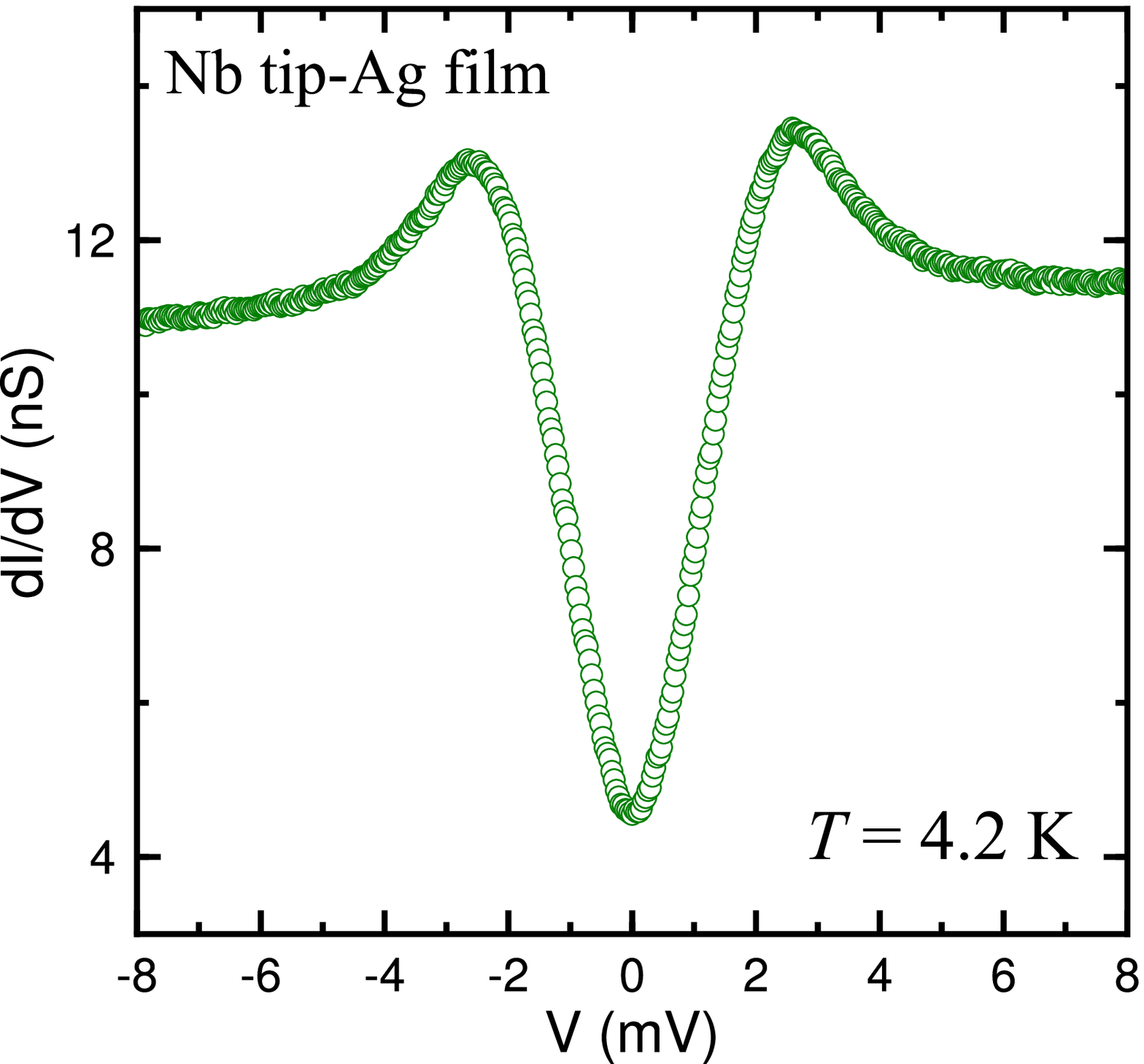}
\caption{\label{fig:NbSTS} STS data taken on a Ag film with a Nb tip at 4.2 K. Feedback settings: sample bias -18 mV and tunneling current 200 pA.}
\end{figure}

For PCS measurements, $dI/dV$  vs. bias voltage was measured using a four-terminal geometry with an AC resistance bridge. For STS measurements, standard two-point AC lock-in technique was used. For PCS, the STM approach method was used to gently bring the tip and sample into contact without crashes. Namely, a piezo stepper motor with a $\sim10$ nm step size was used to move the tip into tunneling range of the sample. The STM feedback was then disengaged, the four-terminal wiring engaged, and the tip gently brought into contact with the sample using the STM piezo.   Nb tips of 99.9\% purity were cut, then cleaned \textit{in situ} through high-voltage field emission on a Ag film. A typical $dI/dV$ spectrum taken by STS with these tips is shown in Figure \ref{fig:NbSTS}. Although the superconducting gap of Nb is $\sim1.5$ meV at 0 K, spectral broadening due to finite quasiparticle lifetime and broadening of the Fermi-Dirac function at 4.2 K causes the coherence peaks in STS data to shift to $\sim\pm$ 2.5 mV.

\newpage

The Bi$_{2}$X$_{3}$ single crystals were grown with the self-flux method \cite{Fisk1989}, using excess Se melt for Bi$_{2}$Se$_{3}$ and excess Te melt for Bi$_{2}$Te$_{3}$. The crystals were cleaved either in air or in a N$_{2}$-filled glovebox. Immediately after cleaving, contacts were made on the freshly exposed $c$-axis surface of the crystals using Ag paint, shortly before being loaded into the STM. Typical atomically-resolved STM topographs of Bi$_{2}$Se$_{3}$ and Bi$_{2}$Te$_{3}$ crystals are shown in the insets of Figures 3 and 4 in the main text, respectively. 

\section{S3. Current flow through BBS}

\begin{figure}[ht]
\includegraphics[width=0.65\textwidth]{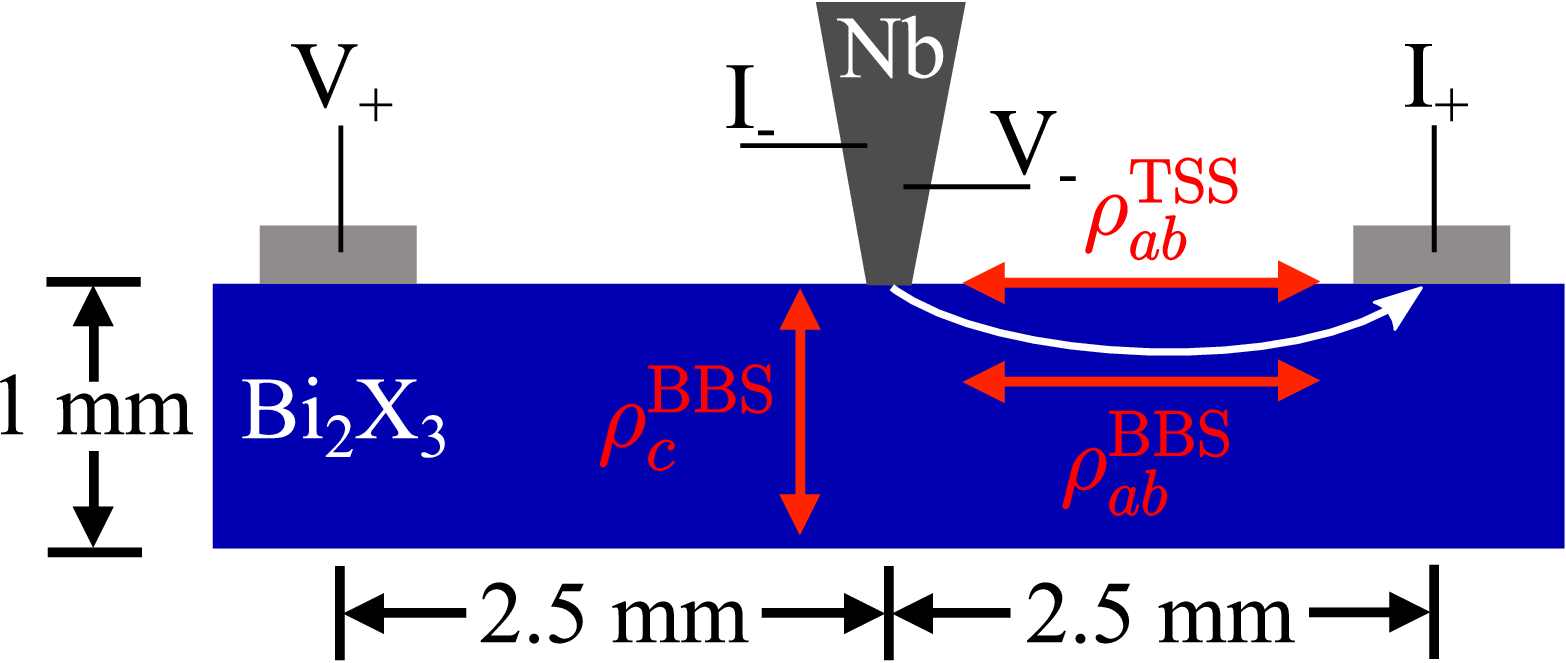}
\caption{\label{fig:current_flow} Schematic of our experimental geometry. The BBS shunt much of the current, as indicated by the white arrow.}
\end{figure}

Figure \ref{fig:current_flow} shows a schematic of the contact geometry used in our experiment. It is important to note that in this geometry, a significant amount of current flows via BBS along the $c$-axis of the crystal. This predominance of BBS is due to the large difference in resistance between TSS and BBS, as detailed below. 

The sheet resistance corresponding to the TSS in our crystals (see Figure \ref{fig:current_flow}) can be estimated as $R_{\Box}=3.69\,k\Omega$ from Ref. [\onlinecite{Checkelsky2011}], which used electrostatic gating and Ca doping of Bi$_{2}$Se$_{3}$ crystals to shift the Fermi level into the bulk band gap. For non-ideal Bi$_{2}$Se$_{3}$ crystals that have dopings similar to ours, the $ab$-plane and $c$-axis resistivities corresponding to the BBS are $\rho_{ab}^{\mathrm{BBS}}=0.25\,\mathrm{m}\Omega\mathrm{cm}$ \cite{Analytis2010} and $\rho_{c}^{\mathrm{BBS}}=1.1\,\mathrm{m}\Omega\mathrm{cm}$ \cite{Gobrecht1969}.

Using the above values, we find that the resistance corresponding to our crystals' BBS is 6 orders of magnitude smaller than that of the TSS: the dimensions of our crystals are $\sim$ 1 x 5 x 5 mm, so the $ab$-plane and $c$-axis BBS resistances for Bi$_{2}$Se$_{3}$ can be estimated as $R_{ab}^{\mathrm{BBS}}=\left(0.25\,\mathrm{m}\Omega\mathrm{cm}\times0.25\,\mathrm{cm}\right)/(0.1\,\mathrm{cm}\times0.5\,\mathrm{cm})=1.25\,\mathrm{m}\Omega$, and $R_{c}^{\mathrm{BBS}}=\left(1.1\,\mathrm{m}\Omega\mathrm{cm}\times0.1\,\mathrm{cm}\times2\right)/(0.1\,\mathrm{cm}\times0.1\,\mathrm{cm})=22\,\mathrm{m}\Omega$. Thus, the BBS shunt much of the current.

\section{S4. Background conductance normalization}

\begin{figure}[ht]
\includegraphics[width=0.4\textwidth]{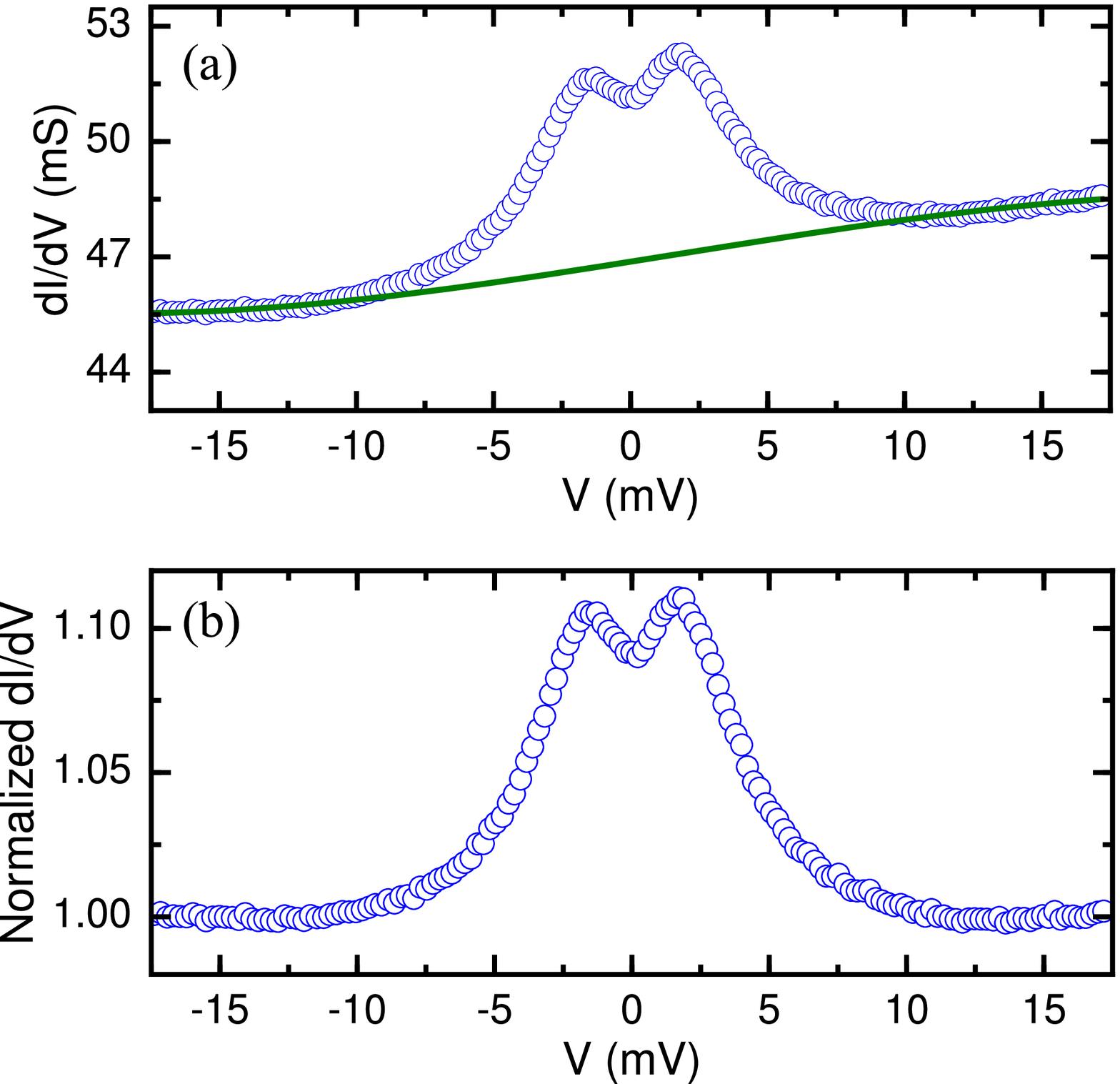}
\caption{\label{fig:fitting}(a) PCS data from Figure 1 in the main text (open circles) and a polynomial fit (solid line) to the conductance background above $\pm$ 10 mV. (b) Normalized data, obtained by dividing the $dI/dV$ data by the polynomial fit.}
\end{figure}

Our PCS data can be normalized by dividing out the background conductance, after interpolating the latter from outside $\sim\pm$ 10 mV using a polynomial fit. An example of this procedure is shown in Figure \ref{fig:fitting}, using Bi$_{2}$Te$_{3}$ PCS data from Figure 1 in the main text. The normalization largely removes the spectral asymmetry, as shown in panel (b).

\section{S5. BTK modeling of $\text{Bi}_{2}\text{X}_{3}/\text{Nb}$ junctions}

\begin{table}[bh]
\begin{center}
\begin{tabular}{|c|c|c|c|c|}
\hline
 & $v_{F}^{c\text{-axis}}$ $\left(10^{5}\,\text{m/s}\right)$ & $v_{F}^{ab\text{-plane}}$ $\left(10^{5}\,\text{m/s}\right)$ & $k_{F}^{c\text{-axis}}$ $\left(\text{nm}^{-1}\right)$ & $k_{F}^{ab\text{-plane}}$ $\left(\text{nm}^{-1}\right)$ \\ \hline
Nb & 2.7 & 2.7 & 4.2 & 4.2 \\ \hline
TSS in Bi$_{2}$Se$_{3}$ & -- & 5.0 & -- & 1.0 \\ \hline
BBS in Bi$_{2}$Se$_{3}$ & 3.2 & 6.7 & 1.5 & 0.7 \\ \hline
TSS in Bi$_{2}$Te$_{3}$ & -- & 4.0 & -- & 1.5 \\ \hline
BBS in Bi$_{2}$Te$_{3}$ & 7.1 & 1.2 & 0.6 & 0.3 \\ 
\hline
\end{tabular}
\end{center}
\caption{\label{tab:Fermi_values} Fermi velocities $v_F$ and wavevectors $k_F$ used in our BTK models. Values are taken from  Refs. [\onlinecite{Analytis2010,Gobrecht1969,Crabtree1987,Xia2009,Chen2009,Koehler1976,Qu2010}]. The Nb values are averages over $k$-space, to account for our polycrystalline Nb tip.}
\end{table}

For the single-interface BTK model in the main text, we apply the formulas in the extended BTK model derived by Mortensen et al. \cite{Mortensen1999}. In this model, the authors derive an angle-dependent $Z(\theta)$:

\begin{align*}
Z(\theta)=\sqrt{\Gamma(\theta)\left(\frac{Z_b}{\cos\theta}\right)^2+\frac{\left(\Gamma(\theta)r_v-1\right)^2}{4\Gamma(\theta)r_v}},
\label{eq:Z}
\end{align*}

where $\Gamma(\theta)\equiv\cos\theta/\sqrt{1-r_{k}^2\sin^2\theta}$, and $r_k=k^N_F / k^{SC}_F$ is the Fermi wavevector matching term mentioned in the main text. This model was used to generate Figure 5(a) in the main text. In our simulation, the $r_v$ and $r_k$ values were fixed by the $c$-axis $v_F$ and $ab$-plane $k_F$ values in Table \ref{tab:Fermi_values}, respectively. The other parameters in the model---$Z_b$, the superconducting energy gap $\Delta$, and a finite quasiparticle lifetime term $\Gamma$---were used as fitting parameters, while $T$ was fixed at 4.2 K. The values for these fitting parameters were 0.3, 1.5 meV, and 1.1 meV, respectively.  As stated in the main text, this BTK model indicates that TSS contribute negligibly to $c$-axis AR, when BBS are also present.

\begin{figure}[th]
\includegraphics[width=0.35\textwidth]{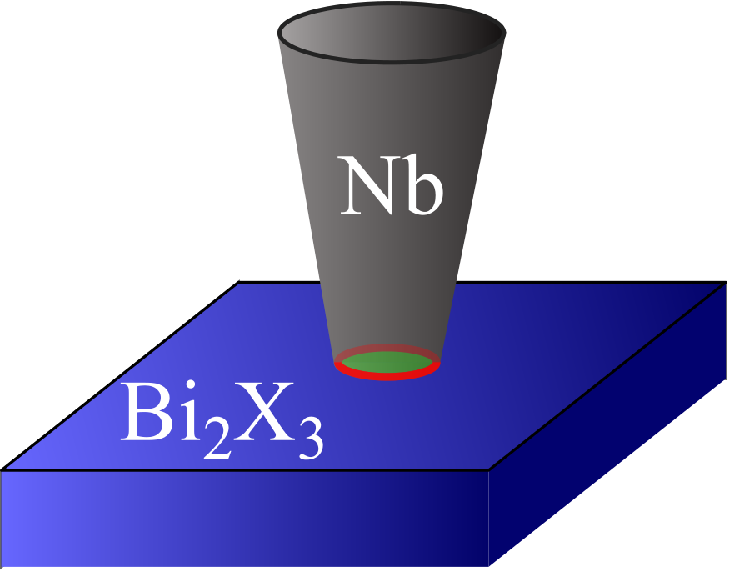}
\caption{\label{fig:BTK_geometries} Illustration of a Bi$_2$X$_3$/Nb junction used for our multi-interface BTK model. The interface for the BBS is a circle of area $\pi a^2$, while the interface for the TSS is a ring of circumference $2\pi a$.}
\end{figure}

For our multi-interface BTK model, we assume that a Bi$_{2}$X$_{3}$/Nb junction has both $c$-axis and in-plane interfaces, as shown in Figure \ref{fig:BTK_geometries}. First, we apply the single-interface BTK model to each interface, again using the $k_F$ and $v_F$ values from Table \ref{tab:Fermi_values}, but this time using the $ab$-plane $v_F$ for the TSS. $Z_b$, $\Delta$, and $\Gamma$ were used as fitting parameters to resemble the data in Figure 2(a), and had values of 0.35, 1.5 meV, and 1.35 meV, respectively. $T$ was again fixed at 4.2 K.

Next, to weight the BBS' and TSS' contributions to the total conductance spectrum, we estimate the number of Landauer conduction channels $N_c$ for each interface. For the BBS' $c$-axis interface of area $\pi a^2$, $N_c =\left(k_{F}^{2}\pi a^2 \right)/\left(4\pi\right)= \left(k_{F} a \right)^2 /4$. For the TSS' in-plane interface of circumference $2 \pi a$, $N_c = \left(k_{F}2\pi a\right)/\pi =2 k_{F} a $.

Finally, we apply these weightings to the BTK spectra for each interface. While we find double-hump AR characteristics for both the BBS/Nb and TSS/Nb channels, once again, the BBS predominate over the TSS. Namely, $N_c^{\text{BBS}}\approx5,200$ and $N_c^{\text{TSS}}\approx260$, as plotted in Figure 5(b) in the main text.

\bibliography{./AR_Bi2X3_PE_supp_mat}